# MOLECULAR DESIGN METHOD BASED ON NEW MOLECULAR REPRESENTATION AND VARIATIONAL AUTO-ENCODER

Li Kai, LiNing, Zhang Wei, Gao Ming

D School of Computer Science+Beijing Advanced Innovation Center for Materials Genome Engineering, Beijing Information Science and Technology University, Beijing, China

## ABSTRACT

*Based on the traditional VAE, a novel neural network model is presented, with the latest molecular representation, SELFIES, to improve the effect of generating new molecules. In this model, multi-layer convolutional network and Fisher information are added to the original encoding layer to learn the data characteristics and guide the encoding process, which makes the features of the data hiding layer more aggregated, and integrates the Long Short Term Memory neural network (LSTM) into the decoding layer for better data generation, which effectively solves the degradation phenomenon generated by the encoding layer and decoding layer of the original VAE model. Through experiments on zinc molecular data sets, it is found that the similarity in the new VAE is 8.47% higher than that of the original ones. SELFIES are better at generating a variety of molecules than the traditional molecular representation, SELFIES. Experiments have shown that using SELFIES and the new VAE model presented in this paper can improve the effectiveness of generating new molecules.*

## KEYWORDS

*VAE, Molecular notation, Multilayer convolutional network, Fisher information, LSTM*

## 1. INTRODUCTION

The discovery of new materials is a crucial driver of many recent technological advances, such as clean energy[1] and drug discovery[2]. New materials allow scientific research to advance, leading to the development of compounds and formulations with new uses, lower costs and better properties. The "similarity hypothesis" proposed by Johnson et al[3]. that compounds with similar structures have similar chemical properties and biological activities have greatly facilitated the discovery of new molecules. However, optimization in molecular space is extremely challenging because chemical searched spaces is huge, discrete, and unstructured[4]. By combining computers with chemistry, molecular information can be calculated, processed, stored and searched more quickly.

In order for a computer to process chemical molecules more efficiently and obtain useful information, it is necessary to use a specific notation to represent the molecule as an input or output to the computer. Therefore, it is necessary to set the molecular descriptor as a string of characters and numbers. Each descriptor can represent the structure of the molecule (such as atomic number, bond number, etc.). SMILES[5] (Simplified Molecular Input Line Entry System) successfully solves the problems of molecular representation and recognition in computers by using simple grammar and two-dimensional molecular coding techniques. SMILES is a simple string used for input and representation of molecular reactions. It is based on the principles of molecular graph theory and can describe molecular structures using simple rules. SMILES have since become a standard tool for chemical informatics.

The rise of artificial intelligence and deep learning is a boon to cheminformatics because they can produce powerful probabilistic generation models that, after experimental calculations, can be simulated to efficiently generate new molecules with excellent properties[6]. These models typically use continuous strings, so generating models can work well with SMILES. However, this presents a key problem: a large proportion of the new SMILES strings generated by generating models are invalid. They may not be syntactically valid, i.e. the string cannot be converted to the corresponding molecular diagram; It may violate basic chemical principles. Many researchers have also proposed some solutions, such as adjusting the generation model to handle the inefficiency of the SMILES strings generated[7]. While this solves the problem of generating a specific model, it is not universal. Another approach is to change the definition of the SMILES strings themselves. O'Boyle et al.[8] invented that Deeps SMILES can be used as a direct input to most machine learning models, overcoming most problems with molecular synthesis and thus smoothly converting the resulting strings of legitimate molecular diagrams. However, it does not handle domain-specific semantic constraints. In the last two years, a string-based molecular SELFIES[9] (SELF-referencing Embedded Strings) invented by the Aspuru group can effectively solve this problem.

In order to study whether the new molecular representation SELFIES can be combined with the deep learning model to produce a good result, this article will use this string to explore experimentally. The first chapter introduces the changes of the core points and methods of molecular design in the last 30 years, and introduces the idea of variational automatic encoder. The second chapter describe the characteristics of the new molecular representation SELFIES and discusses the advantages of the new Variational Auto-Encoders designed in this paper in the field of new molecular design. The third chapter introduce the experimental conditions and preparation and discusses the experimental results. The fourth chapter summarize the content of this paper and looks forward to the opportunities and challenges to chemical informatics.

## 2. RELATED WORK

### 2.1. Major Approaches to New Molecular Design

Molecular similarity is a dynamic and developing field, and many researches have been made on molecular similarity. Willett et al.[10] reviewed molecular similarity and pointed out that similarity search is very important to chemical informatics. Sheridan et al.[11] have made a number of experimental demonstrations of molecular similarity. After a long time of experiments, people found that a lot of information can be found from the method and application of molecular similarity. However, when evaluating some molecules, the final results cannot be integrated due to the different representation or calculation methods of the data used, which brings many difficulties for the subsequent research. Therefore, molecular descriptors appear, which can be converted into topological descriptors to represent molecular structures, achieving the purpose of unified representation. SELFIES is a string-based molecular representation that works well for the computational problems posed by the structure of SMILES itself. The Aspuru group proved through experiments that each SELFIES can contain one effective molecule, which is 100% robust. Besides, SELFIES can be independent of the machine learning model so that you can directly use them as input without any modification to the model. They have also created a set of rules, based on SELFIES, to convert the commonly used SMILES strings of SELFIES today, so that the molecule is absolutely valid before it reaches the encoder.

After unifying the descriptors, researchers focus on calculating the similarity between two molecules. There are two commonly used methods. One is Euclidian distance[12], which is an international definition of distance and can represent the true distance between two molecules in chemical space. The other is the Tanimoto coefficient[13], which mainly calculates the similarity between individuals for the symbolic or Boolean measures, so it can be calculated only by giving the SMILES string, which can be calculated in the range of [0,1]. The closer the result is to 1, the more similar the two molecules are. Through the above two molecular similarity calculation methods can be very intuitive to see the molecular activity in chemical space.

The main problem in designing new molecules are how to efficiently sample from large amounts of data[14]. In the early stage, the probability calculation method was generally used to reduce the molecular structure space, but even if the space was reduced, it could not be exhaustive. This method lacks effective exploration and needs to consume a lot of resources for calculation. Later, people optimized the design and calculation of complex molecules by Monte Carlo method[15]. Through random steps, efforts are made to capture interrelated conditions in molecular design to produce effective molecules. But in a continuous chemical space, it is also difficult for either approach to produce the desired optimal molecule, because the resulting "atomic cloud"[16] may not be a local minimum. In the subsequent design, Tang Yuhuan et al.[17] used the method of combining virtual screening and molecular fingerprint to accurately describe the molecular structure. However, the limitation of this method is high, and it is difficult to carry out a large number of calculations in complex chemical space. An ideal molecular design method should have the following two conditions: 1. It can be quickly and efficiently sampled from complex data centers; 2, processing data can do gradient optimization of their ability, so that the new molecules have stability. Therefore, deep learning has been introduced.

## 2.2. Deep Learning

Deep Learning has become a powerful tool to solve the problem of molecular computation in chemical informatics[18]. Deep learning can be used to solve the problem of excessive molecular data. It can generate a model to train a large amount of data and generate similar data, so as to model the data. These models can extract average and significant characteristics of molecules, and the adaptive architecture of the generated models also enables simple training processes based on back propagation. These models can often be represented using low-dimensional data, allowing analogical reasoning with everyday applications such as natural language. Therefore, many researchers put the idea of deep learning into computational chemistry, generating new molecules through models or predicting the properties of existing molecules. Early Deep learning models such as RBM (Restricted Boltzmann Machine)[19] and DBN (Deep Belief Networks)[20] successfully solved the problem of probability distribution in molecular computing. However, with the increase of the degree of abstraction of molecular design, the early generation model is no longer suitable for further calculation of molecular design due to problems such as insufficient flexibility and ease of handling. Therefore, it is necessary to find a deep learning model more suitable for molecular calculation.

Among the more popular deep learning generation models, VAE (Variational Auto-Encoder)[21] receives increasing attention. VAE is a probability generation model, whose main purpose is to generate brand-new data which is very similar to the original data based on the existing data. VAE is a generation model containing potential variables. It consists of three parts: an encoder network of converting the input data onto fixed dimension vectors, a potential space for mapping vectors, and a decoder network for converting vectors into readable data, so as to generate new data not included in the input data. VAE is trained to receive and learn valid molecules through an encoder. The molecules cross potential Spaces and are randomly sampled by a decoder to generate new molecules. VAE can increase the randomness of encoder and combine with punishment mechanism to make all areas of potential space decode effectively. Rafael Gomez-Bombarelli et al.[22] performed experiments such as new molecule generation and attribute prediction using SMILES, a molecular representation based on the QM9 dataset. However, during the experiment, it was found that the SMILES representation, due to its character-for-character nature and internal syntax, may cause the decoder to produce output invalid molecules. To do this, they modified the decoder model to sample the character probability distribution of each position generated by the last layer of the decoder to output multiple SMILES strings of a single potential space. RDKit[23] was then used to verify that the generated SMILES strings were valid. This can screen for effective molecules, but cannot explain the effectiveness of the model. VAE models for molecular design are shown in Figure 1.

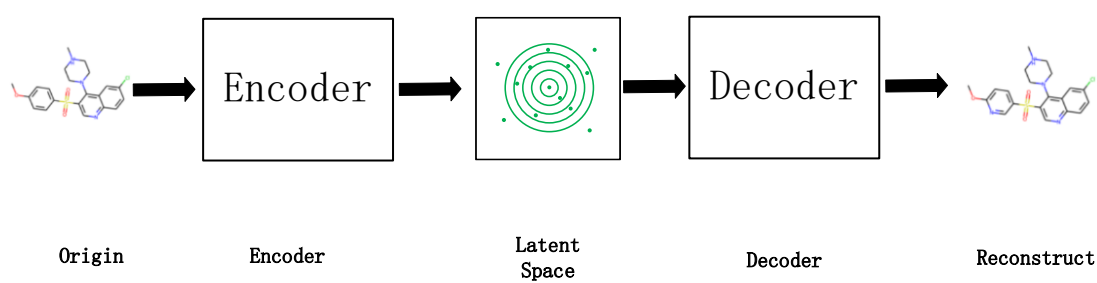

| Origin | Encoder | Latent Space | Decoder | Reconstruct |

Figure 1. Schematic diagram of a variable encoder for molecular design

## 3. NOVEL VARIATIONAL AUTOENCODER FOR IMPROVING THE EFFICIENCY OF GENERATED MOLECULES

The problem facing the VAE model is that only a simple convolutional network is used for encoding and decoding. In this way, although new molecules can be generated, the similarity between the generated new molecules and the original molecules is not stable, that is, the variance may be very large. It is disadvantageous to VAE model to learn potential spatial variables, because in the learning process of the model, the potential variables only obey the prior distribution set in advance, and the meaning behind the data cannot be learned, which will affect the generation effect of decoding and result in degradation. In order to improve the effectiveness of the model, this paper explores the existing encoder and decoder.

### 3.1. VAE Degradation

According to the research of Gammelli, D[24], improving the depth of the model can make the model have better feature expression ability. Improving the likelihood expression function VAE is equivalent to increasing the depth of encoder and decoder. Ideally, enhanced VAE encoders would enable potential Spaces to learn the data. Enhancing VAE decoder can make the result of generation better. Therefore, this paper changes and tests the encoder and decoder respectively, and the results are shown in Figure 2.

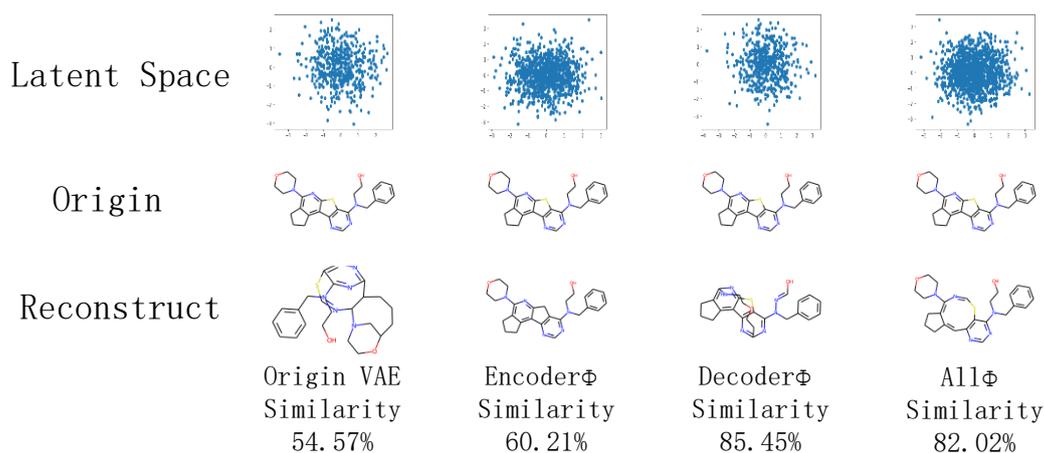

Figure 2.  Comparison of visualized potential Spaces and generated effects when VAE model is enhanced

When VAE model is not changed, the potential Spaces are sparse and the similarity between the generated molecules and the original molecules is poor. With enhanced encoders, the density of potential space increases significantly, This indicates that the potential space are interpretable. Although the effect of generating molecules has been improved, the final effect is still unsatisfactory. After enhancing the decoder, the performance of the potential space is still sparse, the potential space is poor, and the generation effect is obviously improved. However, this may be the result of forcing the decoder to learn data, rather than the excellent data given by the optimized potential space. Finally, when encoder and decoder are enhanced at the same time, the density of potential space is obviously increased, and the generated result is also relatively good. However, it is impossible to judge whether the generated result is better because of the potential space or just because of the enhanced decoder. The above comparison methods show that, for the generation of new molecules, only enhancing the depth of the encoder and decoder cannot effectively explain whether the model capability is improved, and it is easy to lead to degradation. This paper studies the relationship between data information and potential space in order to improve the ability of VAE.

## 3.2. Improve methods to Enhance VAE Ability

The key to improve VAE model ability lies in how to make the encoder produce more useful information, how to smooth the distribution of potential space, and how to effectively learn the data in potential space. For the encoder part, this paper will introduce Fisher Information[25] in information theory, which is a measure of the information contained by random variable X about the unknown parameters of its own random distribution function. Strictly speaking, it is the expected value of the difference or observed information. Suppose that the observed data $X_1, X_2, ..., X_n$ conforms to a probability distribution $f(X;\theta)$, where represents $\theta$ the target parameter, then the likelihood functions can be expressed as:

$$L(X;\theta) = \prod_{i=1}^{n} \frac{\partial \log f(X_i)}{\partial \theta} \qquad (1)$$

In general, the divergence distance presented in formula (1) is positive and convex. Therefore, when the second derivative exists and the likelihood function reaches its optimal value, the first derivative is zero. Fisher's information is represented by $I(\theta)$, and the first-order derivative is represented by $S(X;\theta)$, thus:

$$I(\theta) = E[S(X;\theta)^2] - E[S(X;\theta)]^2 \qquad (2)$$

The mathematical meaning of Fisher information obtained through calculation is as follows: with the increase of the amount of data, the variance between the maximum likelihood function will become larger and larger, which also means that more and more information is mastered. Larger Fisher information means better understanding of the distribution parameters. Intuitively, larger Fisher information means that we can find the true value of the parameter faster and more accurately, and smaller Fisher information means that we need to search for more possible parameters. Therefore, Fisher information can be used to measure the quality of the distribution expressed by the coding layer in VAE to judge whether the encoding result of the encoder is qualified.

For the decoder part, this paper uses LSTM (Long Short Term Memory)[26] as the decoder. The LSTM network can learn the characteristics of the training object, then answer the questions related to this object, and even generate new similar objects. The above characteristics make LSTM neural network a method of chemical informatics, especially in the generation of new molecules has great potential. The new VAE designed in this paper is shown in Figure 3.

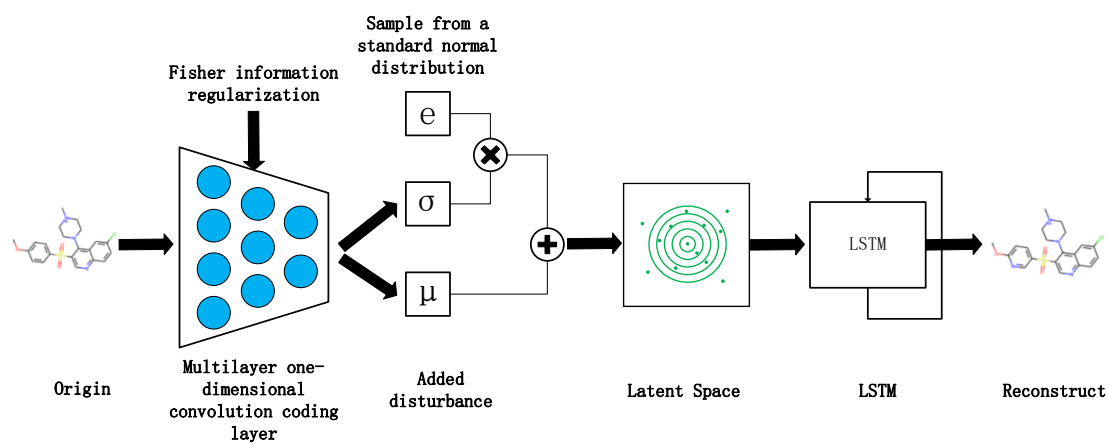

Figure 3. New VAE used to improve the availability of generated molecules

Firstly, a multi-layer one-dimensional convolutional network is used as an encoder, which is used to convert the converted data into a sequence of fixed length of data sets of string type. The purpose of this approach is to treat the process of generating new molecules as string processing. By compressing the string of a vector of specified length, regularized Fisher information is added to the coding layer, and the model actively learns effective sequences, so that the potential space can be better captured and optimized, avoiding the problem of sparse potential space learned by the encoder. After that, LSTM decoding layer is used to learn the data information about potential space and capture the effective data in potential space according to the used data set.

VAE is essentially a generative model. In addition to the invariance of convolutional neural network, over-fitting of input data will inevitably lead to unsatisfactory model generalization ability. As a result, the Vae model is identical with the new one. If many of the generated molecules are identical with the original molecules, it means that the model is failing. At this time, the VAE potential Spaces are not distributed according to the expected idea, that is, the characteristics of the input data are retained and new data are generated. In order to avoid the occurrence of a large number of similar cases, this paper optimizes the data generated by the coding layer. When the original molecule SELFIES is encoded by the multi-layered one-dimensional convolutional network, the potential space can obtain the potential state encoding of all the input molecules. In this case, based on the obtained potential encoding, this paper constructs two layers ($\mu, \sigma$) to be learned. Where, $\mu$ represents the mean value, $\sigma$ represents the variance, and these two layers represent the distribution model of potential space. By sampling e obtained from the standard normal distribution, setting random number seeds, calculating and sampling according to the mean value and variance generated by the encoder, a random disturbed data is constructed. In this way, the encoder can avoid simply reconstructing the original features and improve the effectiveness of the model. As the potential Spaces in VAE model obey normal distribution, the mass of random variables is not near the mean value, but in a ring around the mean value. This means that linear interpolation between two points may pass through a low probability region[27]. This means a variable at some point in the potential space may have a poor effect, but will still be captured by the decoder, resulting in the resulting molecule being of low quality. Therefore, KL divergence should be added in the calculation process to make the model retain the most information in the original data source, and control the decoder to avoid the capture of the low probability region.

## 4. CORRELATION EXPERIMENT

### 4.1. Preparation for Experiment

VAE is adopted for experiments in this paper. In order to simplify model training, SMILES of about 250,000 zinc molecules randomly extracted from the ChEMBL database[28] are used as the data set to measure the effectiveness of deep learning methods in quantum chemistry. This dataset use only the molecular subset of organic atoms (containing only the basic elements H,C,N,O, etc.), contains biologically relevant molecules represented in zinc-associated protonated forms, and is organized into molecular data sets that can be tested quickly. In this paper, the collected data are processed and separated. Among them, 149,673 pieces of data are used as training sets, 49,892 pieces of data are used as verification sets to cooperate with training, and 49,891 pieces of data are used as test sets to test and analysis the results generated by the generation model.

The experiment environment used in this paper is ubuntu1804 and Python version 3.7.0. The experiment adopts the open source advanced framework Pytorch[29]. It is relatively easy about the development and application of deep neural networks, and can be well combined with Python language, making it easier about people to carry out scientific research calculations.

The main steps of the experiment are as follows: Converting the existing SMILES dataset into SELFIES, encoding it into the encoder through One-Hot, converting each molecule into a vector in the potential space. Then, the decoder captures a point in the region of the potential space, converting the vector from that point in a corresponding string, and finally, using the designed program for verification and analysis. Generate the desired result graph. After repeated experimental verification and trial and error, batch_size was set to 256, a total of 500 rounds of training was set to 0.0001, the learning rate were set to 0.0001, and the model architecture as shown in Figure 4 was selected to use GPU for training. The results obtained were discussed in the next section.

```
-----------------------------------------------------------------------
Layer (type)                        Output Shape              Param #
=======================================================================
Conv1d-1                           [-1, 24, 72]               28,536
Conv1d-2                           [-1, 24, 72]                6,360
Conv1d-3                           [-1, 24, 72]                6,360
Linear-4                             [-1, 256]               442,624
Linear-5                             [-1, 256]               442,624
LSTM-6                           [-1, 108, 792]             9,611,712
Linear-7                         [-1, 108, 72]                57,096
=======================================================================
Total params: 10,595,312
Trainable params: 10,595,312
Non-trainable params: 0
-----------------------------------------------------------------------
Input size (MB): 0.029663
Forward/backward pass size (MB): 0.755371
Params size (MB): 40.417908
Estimated Total Size (MB): 41.202942
-----------------------------------------------------------------------
```

Figure 4. Model architecture used in this experiment

## 4.2. Result

Using SELFIES as the input, we won't use as many SELFIES as SMIELS for direct observation, so we can use the trained VAE model to generate new SELFIES to convert them to SMILES for better observation and subsequent experimental verification. As the new molecular string generated is sampled from the last layer of decoder, the decoding process of VAE model is uncertain, which means that the decoding process of a single vector is random in the potential space. So, when a molecule is captured by the decoder and enters the decoding process, other SELFIES can be re-encoded into the potential space to wait for the next capture by the decoder. The region captured most frequently by the decoder shows that the molecule in this region is the most similar to the original one, which can be known by calculation. The captured molecule is the closest to the Euclidean distance from the original molecule, and the distance between other molecules in the region and the original molecule is proportional to the distance between other molecules and the captured molecule, that is, the farther away from the captured molecule, the lower the similarity. Figure 5 shows the Euclidean distance between the molecules in the region and the captured molecules when a molecule is captured. As can be seen from Figure 5, the closer they are to the captured molecules, the more similar their molecular diagrams will be to the captured molecules. When the similarity is relatively high, the main structure of the generated molecule will not change, but small changes will occur in the secondary structure, such as the addition of new elements or ring disintegration.

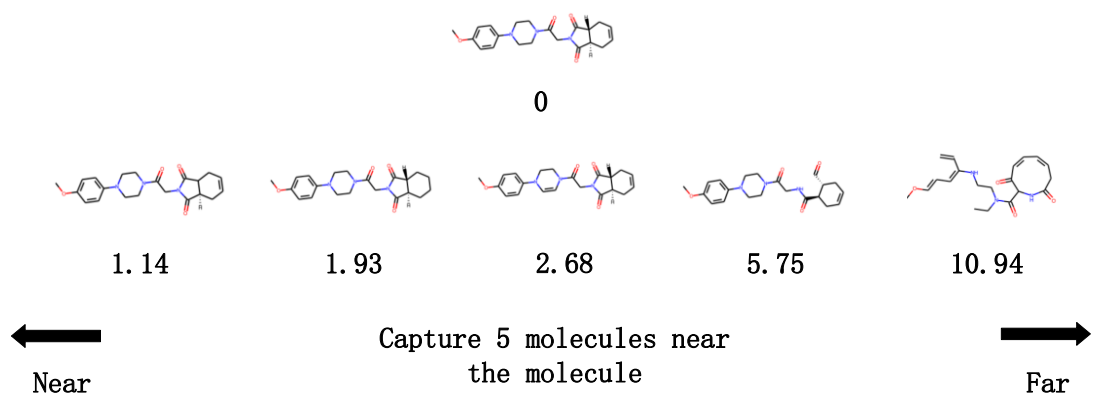

Capture 5 molecules near
the molecule

← Near                                          Far →

Figure 5. Euclidean distance between captured molecules and nearby molecules in potential space

To test the effectiveness of the model, we used SELFIES as input in the model in this article, followed by SELFIES in R Gomez-Barbarella et al.[22] (R Group) for training. After the two models were trained, 100 molecules were randomly selected from the test set to generate and calculate new molecules. It was found that among the 100 generated molecules, the generated molecules were not identical with the original molecules in the two models, and the generated molecules were 100% effective, which was also consistent with the conclusion obtained by Aspuru group. Zhou Wei[30] believed that when the Tanimoto coefficient of two molecules was above 70%, the activity and properties of two molecules were valuable to study. The average similarity of the generated molecules calculated by the R group model is about 70.26%. The average similarity of the generated molecules calculated by this model is about 78.73%. The frequency of the overall similarity coefficient is shown in Figure 6. It can be seen that the similarity span of molecules generated by the R group model is large, and many of the generated molecules have low similarity, and the number of molecules with research value is small; Most of the molecules generated by this model have a similarity interval of [60%, 90%]. The number of low similarity molecules is very small, and there are many molecules with research value. The test data show that the model in this paper shows good performance in terms of the similarity of the molecules generated from the zinc molecular data set.

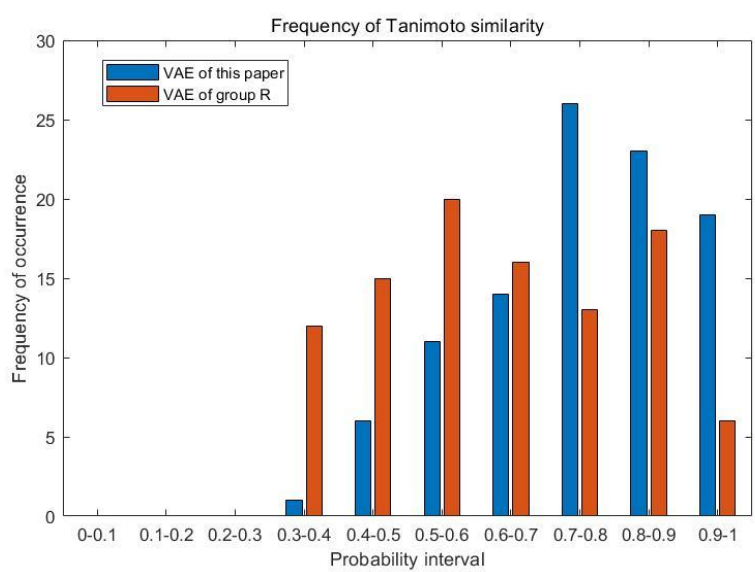

Figure 6. The molecule generated by the two models is similar to the original molecule

In addition to the similarity shown in formation of a single new molecule, the molecular diversity shown in the formation of multiple new molecules is also important because it reflects the density of molecules in the underlying space[31]. The denser the molecules encoded in the potential space, the more kinds of molecules can be decoded, and the better the chemical space can be explored. To verify that SELFIES can increase the molecular density of potential Spaces, we'll use SELFIES and SMILES for experimental verification in this article. After training the data of both molecular representations using the models presented in this paper, a molecule was randomly selected from the test set and 200 samples were taken for each model respectively. The number of valid molecules was also calculated for the model using SMILES, after which similarity was sorted and statistical calculations were performed. We found that the model trained with SELFIES produced 77.5% of diverse molecules, while the model trained with SMILES produced 40.59% of diverse molecules, as shown in Figure 7. The SMILES curve often appears parallel to the horizontal axis, which indicates that the molecules in this part are the same. We shot a few times in parallel, which means that most of the molecules are different. Calculations have shown that using SELFIES can dramatically increase the diversity of molecules.

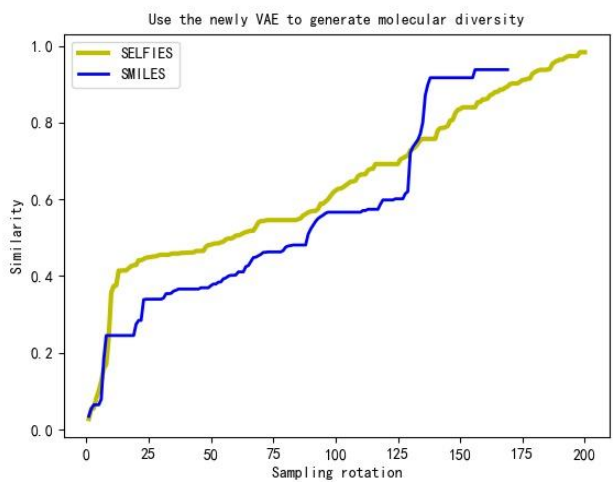

Figure 7. Diversity of generative models trained with different molecular representations

## 5. CONCLUSIONS

The mentioned above experiments show that the combination of one-dimensional Convolutional Neural Network and LSTM neural network can achieve better performance in molecular generation, and the generated results are novel and diverse, with good performance and research space. Using LSTM as a decoder allows you to define the probability distribution of all possible characters at each position of the string, which improves the accuracy of generating the string. The application of SELFIES to the generated model can also provide good reconstruction fidelity and the ability to capture the features of the molecular training set, with 100% effectiveness and the ability to increase the molecular density in the potential space, which can play an important role in reverse design technology in computational chemistry[32]. In the next step, consider changing the input method. For example, by using an existing molecular fingerprint method (such as ECFP[33]), you can directly build a graphic encoder based on SELFIES. Even though the Selfies used in this article is a powerful one, you can save a lot of work by simply putting selfies into the model for training. And directly output molecular diagram, intuitive effect will be better. However, it is difficult to build a neural network that can output arbitrary graphs.

In the training process, due to conditions, only one GPU was used for training, and in order to achieve high quality results, the speed and performance were not satisfactory. Since the data set used in the experiment in this paper only contains SMILES strings, the addition of molecular attribute data will certainly affect the setting of parameters and the complexity of training, and the speed of training will decrease significantly. Therefore, one of the subsequent works of this paper is to improve the training speed, adopt the way of multi-threading, multi-process and multi-GPU combination, modify the training mode of the variational automatic encoder based on text molecular coding, so as to achieve the purpose of high quality and high performance.

## ACKNOWLEDGEMENTS

I would like to express my heartfelt thanks to all those who have helped me in this paper. First of all, I would like to express my heartfelt thanks to my two tutors, Mr. Li Ning and Mr. Zhang Wei, whose advice and encouragement have given me more insight into these translation studies. It is my great pleasure to study under their guidance and guidance. Moreover, it is my pleasure to benefit from their personality and diligence, which I will cherish all my life. I can't thank them enough. I am also very grateful to Mr. Gao Ming who helped me and accompanied me in the process of preparing this paper. I also want to thank my family for their love and unwavering support. Finally, I am very grateful to all of you who took the time to read this paper and gave me many suggestions, which will help me in my future study.

**AUTHORS**

**Li Kai**, master student, main research field is Material Genetic engineering and machine learning, mobile number: 13521099187, correspondence address: Beijing Information Science and Technology University, No. 35, Middle North Fourth Ring Road, Yayuncun Street, Chaoyang District, Beijing, Postcode: 100101. E-mail: 923436067@qq.com

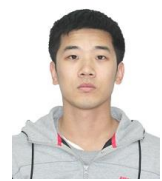

**Li Ning**, Professor, main research field is Document Information processing, XML and Information technology standardization. E-mail: lining@bistu.edu.cn

**Zhang Wei**, corresponding author, professor, research interest covers network and data security, cryptography application technology, software and hardware co-design. E-mail: zhwei@bistu.edu.cn

**Gao Ming**, experimental researcher, main research field: Artificial intelligence, data security, cloud computing. E-mail: gm@bistu.edu.cn